\documentstyle[12 pt,epsf]{article}
\begin{document}

\title{\bf{Equivalence of Minkowski and Euclidean Field Theory Solutions}}
\author{Khin Maung Maung $^{a)}$ \thanks{maung@jlab.org} \, Charles A. Hill
$^{a)}$ \thanks{charles.hill@hamptonu.edu} \, 
Michael T. Hill $^{a)}$\thanks{michael.hill@hamptonu.edu} \cr \, and
George DeRise $^{a,b)}$\thanks{gderise@cox.net}
       \\
       \\ $^{a)}$ Department of Physics, Hampton University
       \\ Hampton, VA 23668
       \\
       \\  $^{b)}$ Department of Mathematics, 
       \\ Thomas Nelson Community College, 
	\\Hampton, VA 23670}

\maketitle

\begin{abstract}

\noindent
We consider the correspondence between solutions 
of non-gravitational field theories formulated in Euclidean space-time and 
Minkowski space-time.  Infinitely many ``Euclidean'' spaces 
can be obtained from $M^4$ via a group of transformations 
in which the Wick rotation is a special case.  We then discuss 
how the solutions of gauge field theories formulated in 
these ``Euclidean'' spaces have a one-to-one correspondence 
with the solutions of field theories formulated in 
Minkowski space-time, provided we avoid the one-point 
compactification into $S^4 = {E^4\bigcup\infty}$.
\end{abstract}

To solve technical problems encountered in phenomenological 
and theoretical applications of non-gravitational field theories formulated in 
flat Minkowski
space-time with Lorentz metric ($+---$) it is often useful to transform to the 
four-dimensional Euclidean space-time with metric ($++++$).
Solutions are then found in the Euclidean space-time.  We 
discuss the problem of determining whether or 
not these solutions have corresponding solutions in Minkowski
space-time.

Normally, Euclidean space-time is reached via the Wick 
rotation, which is a special case of the more general 
transformation given by $\tau= \alpha_1 t + \alpha_2$,
where $\alpha_1 , \alpha_2~\epsilon~ C$.
Each transformation is an element of the group of 
conformal transformations of the plane onto itself.  To 
obtain the so-called generalized Wick rotation, we take 
$\alpha_1 = e^{i\pi/2} ~{\rm and} ~\alpha_2 = 0$. 

Our intention is to demonstrate that field theory 
solutions in the Euclidean space-time obtained above have 
a one-to-one correspondence with solutions in Minkowski 
space-time provided that the conditions imposed at infinity
in the Euclidean space-time do not amount to a one-point
compactification into $S^4 = {E^4\bigcup\infty}$.  
Let 
the gauge theory in a Euclidean or Minkowski space-time manifold be described by
a principal bundle
($B$,$P$,$\pi$,$F$,$G$).  $B$ is the {\it base space}, which 
is $E^4$ for theories formulated in Euclidean space-time and $M^4$
for theories formulated in Minkowski space-time.
$P$ is the {\it total space} over 
$B$ and $F$ is a {\it typical fiber} of $P$.  The
projection which maps a fiber $F$ from $P$ to a
point in $B$ is $\pi$.  $G$ is the {\it gauge group}, which
is always topologically equivalent to the typical fiber $F$ in a
principal bundle.  The fiber over a point $p$ in the 
Euclidean base space contains solutions in every possible gauge
at point $p$.  A single point on a fiber represents a solution
in a particular gauge.  Different gauges can be reached by
operating on a point in the fiber by an element in gauge
group $G$.  For a curve in $B$ the
solutions of the theory in a chosen gauge will be represented
by single points on each of the fibers above the curve.  The locus of these
points is a cross-section.

\begin{figure}[htbp]
\centerline{\epsfxsize=4.5in\epsfbox{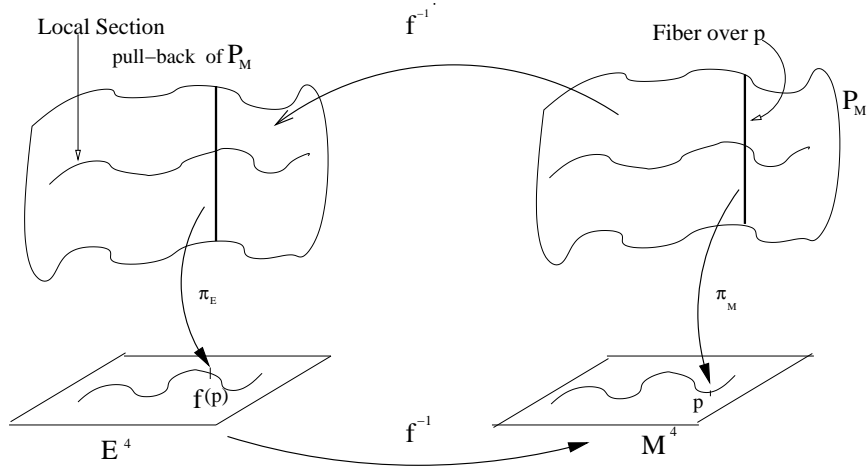}}
\caption{A Wick rotation $f$ maps $M^4$ to $E^4$.  A field theory
in $M^4$ is described by {\it total space} $P_M$.  The corresponding
field theory in $E^4$ is contained in a pullback bundle,
which carries the same {\it typical fiber} as $P_M$. Therefore the
two principal bundles have the same gauge group.  The solutions for
each {\it base space} in a chosen gauge form a section. \label{inter}}
\end{figure}

Consider the principal fiber bundle ($M^4$, $P_M$, $\pi_M$,
$F_M$, $G_M$) describing a Minkowski field theory.  Because 
there is a homeomorphic mapping $f$ (Wick rotation) 
between Euclidean 
space-time $E^4$ and the Minkowski base space $M^4$, 
a unique mapping from $P_M$ to a new space $P_E$ exists.
The induced pullback bundle $P_E$ forms a total 
space over $E^4$ with the same typical fiber as $P_M$.
 The total spaces
$P_E$ and $P_M$ are topologically equivalent and 
the cross-sections are homeomorphic.\cite{naka} 
Now, the cross-section of the pullback bundle $P_E$ is
the pullback of the cross-section of $P_M$.  Therefore
the  field theory solutions have a one-to-one correspondence.
We also note that the base spaces $E^4$ and $M^4$ 
are contractable spaces so the 
two principal fiber bundles formed are {\it trivial}
and hence global cross-sections exist.
 
	In some formal and phenomenological applications 
of Euclidean field theories, assumptions have to be made 
about the asymptotic behavior of the field. In order to achieve
finite action for theories written in $E^4$, appropriate conditions are imposed 
at infinity $($i.e. $r\equiv \sqrt{\sum_{i=1}^4x_i^4} \rightarrow\infty)$.  
If these conditions are equivalent to the one-point compactification
$S^4 = {E^4\bigcup\infty}$, then there no longer exits a  homeomorphism
between the base spaces, i.e. there is 
no mapping between $S^4$ and $M^4$
unless a pole is removed from $S^4$. Therefore the total space over
$S^4$ is not the pullback of $P_M$.  $S^4$ is not
contractable and the total space over $S^4$ is not {\it trivial}.
Furthermore, Singer\cite{singer}
showed that the principal bundle with base space $S^4$ and a compact, nonabelian Lie group
as structure group admits no global sections.  A gauge cannot be
chosen that gives a continuous section everywhere in the total space
above $S^4$.  This problem is known as the Gribov Ambiguity\cite{gribby}.  
Since $P_M$ admits global sections, 
there is no homeomorphism between $P_M$ and the total space over $S^4$.   
This implies that there is no one-to-one
correspondence between the field theory solutions of $M^4$ and $S^4$.  Subsequent compactification
of $M^4$ does not preserve the correspondence since a compactification
of $M^4$ yields a topologically different space $S^1\times S^3$ with 
a flat Lorentz metric\cite{pen}.  This lack of correspondence is seen in nonabelian theories over
$E^4\bigcup\infty$.  Such theories find self-dual and
anti-self-dual solutions satisfying $F = \pm^*F$, where $F$ is the curvature
two-form and $^*F$ is the dual.  In $M^4$, where this relation becomes $^*F = \pm iF$, no 
solution corresponding to a compact Lie group can be found\cite{nash}.

\section*{Acknowledgments}
We would like to thank Peter C. Tandy for encouraging us to look
into this problem. KMM would like to acknowledge the support
of National Aeronautics and Space Adminsnistration  grant NCC-1-251.

\end{document}